\documentclass[pdflatex,sn-mathphys-num]{sn-jnl}

\usepackage{graphicx}%
\usepackage{multirow}%
\usepackage{amsmath,amssymb,amsfonts}%
\usepackage{amsthm}%
\usepackage{mathrsfs}%
\usepackage[title]{appendix}%
\usepackage{xcolor}%
\usepackage{textcomp}%
\usepackage{manyfoot}%
\usepackage{booktabs}%
\usepackage{algorithm}%
\usepackage{algorithmicx}%
\usepackage{algpseudocode}%
\usepackage{listings}%
\usepackage{fvextra}
\usepackage{xurl}
\usepackage{pdfpages}

\theoremstyle{thmstyleone}%
\theoremstyle{thmstyletwo}%

\theoremstyle{thmstylethree}%

\begin{document}

\title[Article Title]{Agentic Literature Extraction Augments Targeted Protein Degradation Databases}


\author[1]{\fnm{Yaochen} \sur{Rao}}\email{yaochenr@chalmers.se}
\author[1]{\fnm{Farzaneh} \sur{Jalalypour}}\email{farjal@chalmers.se}
\author[2]{\fnm{N. M. Anoop} \sur{Krishnan}}\email{krishnan@iitd.ac.in}
\author*[1,3]{\fnm{Roc\'io} \sur{Mercado}}\email{rocio.mercado@chalmers.se}

\affil*[1]{\orgdiv{Department of Computer Science and Engineering}, \orgname{Chalmers University of Technology and University of Gothenburg}, \orgaddress{\city{Gothenburg}, \postcode{412 96}, \country{Sweden}}}

\affil[2]{\orgdiv{Yardi School of Artificial Intelligence}, \orgname{Indian Institute of Technology Delhi}, \orgaddress{\city{New Delhi}, \postcode{110016}, \country{India}}}

\affil[3]{\orgname{Science for Life Laboratory (SciLifeLab)}, \orgaddress{\city{Gothenburg}, \country{Sweden}}}


\abstract{Predictive models in medicinal chemistry depend on structured assay data locked in primary publications. This bottleneck is acute in targeted protein degradation (TPD), where each assay record must combine compound identity, degradation target, recruiter, assay context, and endpoint values reported across sections, tables, and supplementary files. Existing TPD databases are manually curated and often lack assay context. We present an expert-in-the-loop large language model workflow in which a cross-validated prompt-refinement module adapts extraction instructions from scarce expert annotations. With only seven annotated molecular glue publications, the workflow achieved record-level $F_1 = 0.98$ and transferred to proteolysis-targeting chimeras (PROTACs) by terminology substitution alone ($F_1 \geq 0.93$). The resulting database, AutoTPD, contained 81\% and 92\% more molecular glue and PROTAC records than standardized reference datasets; expert audits found 92/100 and 33/40 new records to be correct. Applying the workflow to eligible PubMed Central open-access literature added 4,483 records from 598 publications, yielding AutoTPD+, a public, versioned database of 7,622 TPD records, including degradation assay measurements, designed for monthly updates. In a controlled surrogate model comparison, PROTAC degradation activity models trained on AutoTPD+ saw a 8.9--18.2\% error reduction relative to models trained on the original PROTAC-DB data.}




\maketitle

The biomedical literature is growing at an unprecedented rate, with $>$1.5M articles published annually in PubMed alone~\cite{gonzalez2024landscape}. For data-driven drug discovery, the bottleneck is no longer a lack of information but the difficulty of turning published results into structured, machine-readable data~\cite{bender2021artificial}. This problem is particularly evident in targeted protein degradation (TPD), a rapidly expanding class of therapeutic modalities in which molecules like proteolysis-targeting chimeras (PROTACs) and molecular glues (MGs) redirect the cell's protein disposal machinery to selectively eliminate disease-causing proteins, thereby extending drug discovery beyond conventional inhibition~\cite{zhong2024targeted, gharbi2024comprehensive} and also extending the scope of druggable targets. The TPD literature contains thousands of degradation assay measurements scattered more than a thousand of publications, but the leading public resources all rely on manual curation and cannot keep pace: PROTAC-DB~\cite{ge2025protac} and MolGlueDB~\cite{wang2026molgluedb} compile assay data from primary publications, while TPDdb~\cite{qin2026tpddb} and PROTAC-PatentDB~\cite{cai2025protac} extend coverage to patent-derived compounds. These resources are updated through periodic releases, whereas the literature they summarize continues to grow: in 2026 alone, our search of PubMed Central (PMC) identified more than 200 open-access TPD articles reporting quantitative degradation data, roughly 20 per month.

Across these resources, manual curation prioritizes prominent potency measurements, such as the half-maximal degradation concentration (DC$_{50}$) and the maximal degradation (D$_{\max}$), whereas contextual details such as assay type, cell line, treatment duration and concentration are often missed. This context is important because TPD degradation measurements are inherently condition dependent. Existing resources may also emphasize representative or highly active compounds, leaving weaker or inactive measurements under-represented despite their potential value for downstream analysis and modeling \cite{ribes2024corrigendum}. 

Automated extraction from scientific literature has progressed from rule-based and named-entity-recognition pipelines~\cite{mavracic2021chemdataextractor,isazawa2022single} and optical chemical structure recognition~\cite{rajan2023decimer} to LLM-based methods that operate on full text papers without task specific fine-tuning~\cite{dagdelen2024structured,polak2024extracting}. Recent applications include domain-adapted models and extraction pipelines in materials and polymer science~\cite{gupta2022matscibert,circi2024well,gupta2024data, kunchapu2025polymetrix}, as well as multimodal systems that combine language and vision models~\cite{lederbauer2025lemat,schilling2025text}. Recent work has moved from static datasets to continuously maintained resources, as in PERLA, an autonomous living database for perovskite photovoltaics~\cite{shabih2026autonomouslivingdatabaseperovskite}. In parallel, methods such as GEPA, DSPy and LLM-as-judge or LLM-as-optimizer frameworks provide related strategies for prompt refinement and evaluation~\cite{agrawal2025gepareflectivepromptevolution,khattab2024dspy,rios-garcia2025llmjudge}.

TPD assay records nevertheless pose challenges beyond those addressed by generic extraction pipelines. Each record must combine compound identity, degradation target, E3 ligase recruiter, assay context and endpoint values, even when these fields are reported in different sections, tables or supplementary files. Extraction is further complicated by inconsistent compound identifiers, such as internal numbers, common names and systematic names, and by assay context that is incomplete, implicit or separated from the reported potency values. Despite the growing use of deep learning in degrader design~\cite{gharbi2024comprehensive,dunlop2025predicting,ribes2026protac}, no systematic LLM-based extraction effort has, to our knowledge, been developed for this literature. More broadly, it remains unclear how LLM-based extraction systems can be adapted to a specialized curation task when only a handful of expert-annotated documents are available, how their outputs should be validated against curated resources that are themselves incomplete, and whether such a system can be run on newly published papers without further annotation, so that the resulting database is kept up to date rather than released as a one-off snapshot. TPD therefore provides a demanding test case for AI-assisted curation more generally.

Here we develop an expert-in-the-loop workflow for TPD database curation that extracts structured assay records from full-text articles and supplementary materials, verifies them against source sentences and standardizes chemical and biological entities. A cross-validated agentic prompt optimization module (CAPO) adapts the extraction instructions from scarce expert annotations under constraints that limit paper-specific overfitting. Using only seven annotated molecular glue publications, the workflow achieves a record-level F1 of 0.98 against expert ground truth, transfers to PROTACs by terminology substitution alone ($F1 \geq 0.93$) and retains its performance across three LLM families. Evaluated through a three-way comparison with curated databases and expert annotations, the resulting AutoTPD benchmark contains 81\% and 92\% more molecular glue and PROTAC records than the corresponding standardized reference datasets and recovers assay context that is largely missing from existing resources. Having validated the workflow, we froze and deployed it across additional eligible PMC open-access literature, extracting 4,483 records from 598 publications and producing AutoTPD+ v1.0 (October 2026), a versioned database containing 7,622 records from 1,008 publications, together with a workflow designed to support monthly updates. In a controlled downstream comparison, PROTAC property prediction models trained on AutoTPD+ yielded 8.9--18.2\% lower held-out error estimates than the same models trained on PROTAC-DB. We release the workflow, prompts, evaluation code, expert annotations and versioned datasets, together with a public website, as resources for TPD research and, more broadly, for AI-assisted scientific curation.

\section*{Results}\label{sec2}

\begin{figure}[ht]
  \centering
  \includegraphics[width=\linewidth]{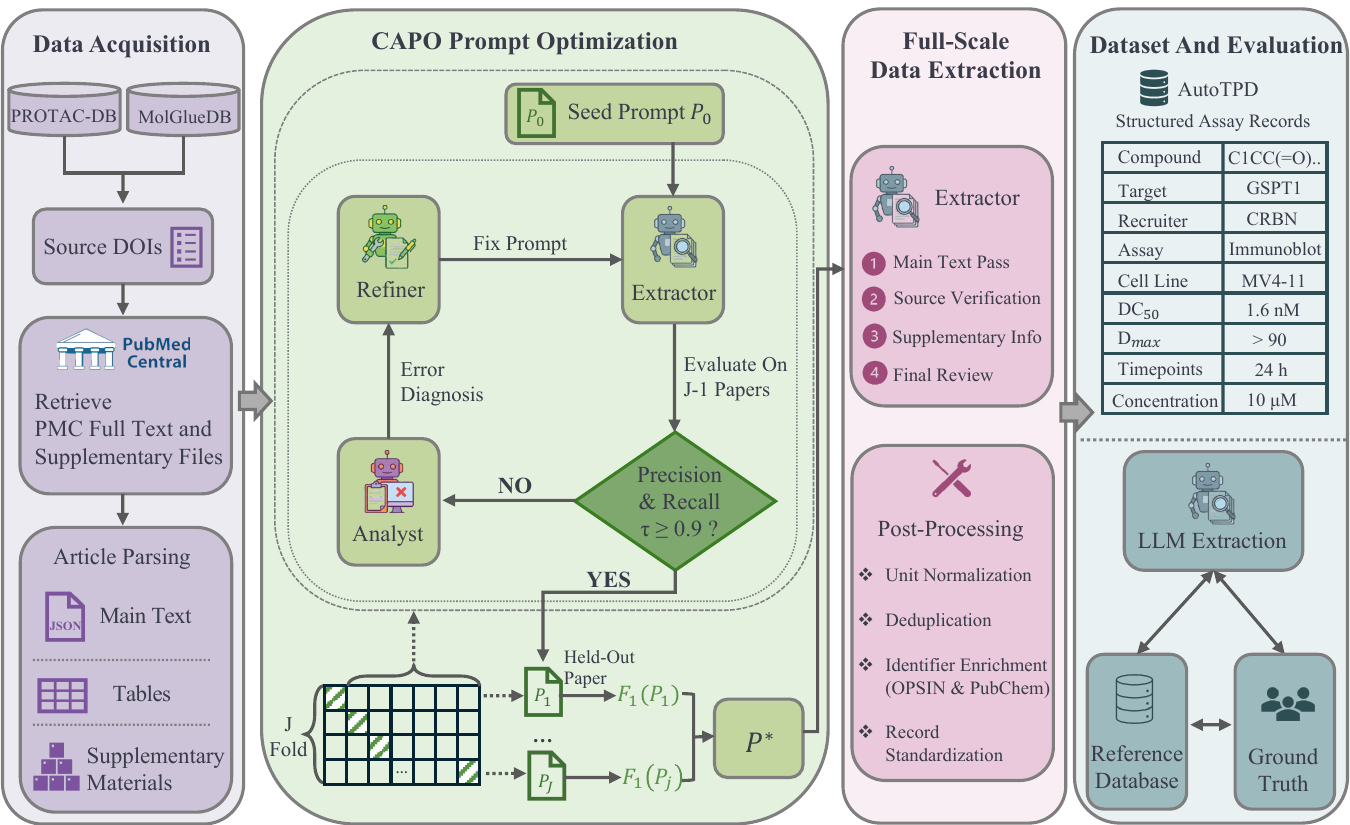}
  
  \caption{Overview of the agentic literature extraction workflow. Source DOIs from PROTAC-DB and MolGlueDB which are available via open-access PMC are extracted as full text and supplementary materials, then parsed into main text, tabular, and supplementary inputs. CAPO optimizes the extraction prompt through leave-one-out assess--diagnose--refine loops and selects a Champion Prompt ($P^*$) based on held-out performance. The frozen prompt is applied at scale, and the resulting records are post-processed to produce AutoTPD for comparison with reference databases and ground truth.}
  \label{fig:pipeline_overview}
\end{figure}

\subsection*{Overview of the AutoTPD curation and evaluation framework}

We developed an expert-in-the-loop workflow that converts full-text articles and supplementary materials into standardized TPD assay records (Fig.~\ref{fig:pipeline_overview}). Open-access articles sourced from PROTAC-DB and MolGlueDB were retrieved from PubMed Central and parsed into main-text, tabular and supplementary inputs. Because expert-annotated papers are scarce, we used Cross-validated Agentic Prompt Optimization (CAPO) to adapt the extraction prompt from a small annotation set. CAPO comprises three agents with distinct roles: an Extractor that identifies assay records and verifies them against the source documents, an Analyst that diagnoses recurring false positives and false negatives, and a Refiner that translates these diagnoses into constrained revisions of the extraction instructions. Once selected, the optimized prompt was frozen and applied to the full molecular-glue and PROTAC cohorts. The extracted records were then normalized, deduplicated, enriched with chemical identifiers and standardized. The workflow was first validated on the database-sourced cohorts to produce AutoTPD, the frozen benchmark dataset used to evaluate extraction accuracy and database augmentation. The validated workflow was then deployed, with the prompt frozen, across additional eligible PMC open-access literature to produce AutoTPD+, a versioned database that integrates the extracted records with standardized reference database records. The accompanying workflow is designed to support monthly literature updates.

\begin{figure}[h]
  \centering
  \includegraphics[width=\linewidth]{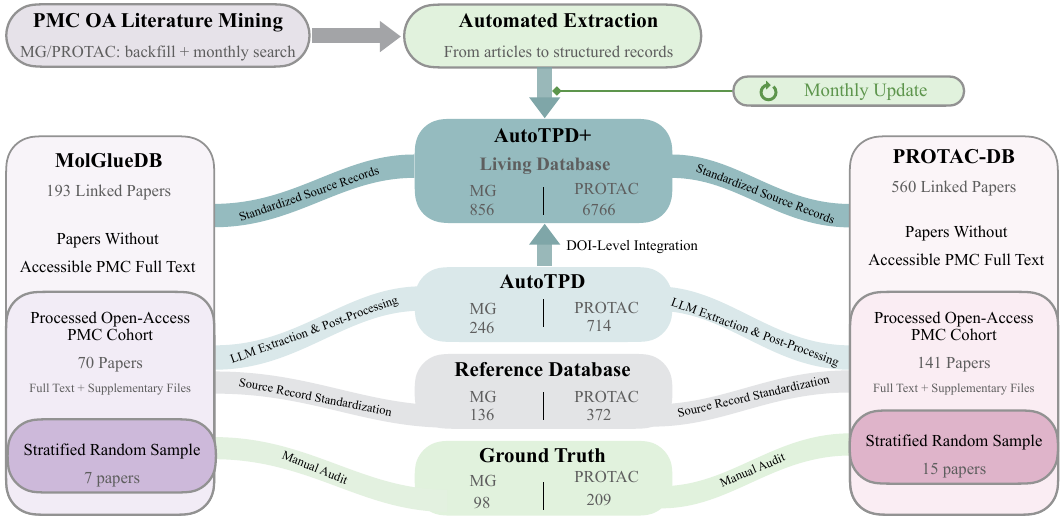}
  \caption{Construction of the AutoTPD benchmark and AutoTPD+. Of the publications linked from MolGlueDB and PROTAC-DB, 70 molecular-glue (MG) and 141 PROTAC papers were processed, yielding 246 and 714 AutoTPD records, respectively; standardized database entries from the same cohorts yielded 136 and 372 reference records. Stratified annotation of 7 MG and 15 PROTAC papers produced 98 and 209 ground-truth records. After benchmark validation, the frozen workflow was deployed across additional eligible PMC open-access literature. DOI-level integration of the extracted records with standardized source-database records produced AutoTPD+ v1.0, containing 856 MG and 6,766 PROTAC records.}
  \label{fig:data_summary}
\end{figure}

CAPO was optimized on the seven expert-annotated molecular-glue papers using document-level leave-one-out cross-validation (LOOCV). In each fold, six papers were used for prompt refinement, whereas the remaining paper was withheld from refinement and used to evaluate the resulting candidate prompt. The candidate with the highest held-out F1 score was selected as the Champion Prompt and subsequently transferred to the PROTAC cohort with only a terminology substitution. We evaluated the complete workflow through a triangular comparison among LLM predictions, standardized records from the source databases and expert-annotated ground truth (Fig.~\ref{fig:pipeline_overview}). The three pairwise comparisons address complementary questions: predictions against ground truth measure extraction accuracy, database records against ground truth audit the coverage of existing resources, and predictions against database records quantify the records and assay context recovered beyond manual curation.

Of the 193 MolGlueDB-linked and 560 PROTAC-DB-linked publications, 70 and 141, respectively, were available as open-access PMC full text and successfully processed (Fig.~\ref{fig:data_summary}). Within these benchmark cohorts, standardization of the reference databases yielded 136 molecular glue and 372 PROTAC assay-level reference records, whereas LLM extraction and post-processing yielded 246 and 714 AutoTPD records. Stratified samples of 7 molecular glue and 15 PROTAC publications were independently annotated, producing 98 and 209 ground-truth records, respectively. AutoTPD served as the frozen benchmark for extraction accuracy and database augmentation analyses. After validating the workflow on these benchmark cohorts, we froze and deployed it across additional eligible PMC open-access literature. DOI-level integration of the resulting records with standardized MolGlueDB and PROTAC-DB records produced AutoTPD+, containing 856 molecular glue and 6,766 PROTAC records in its v1.0 release.

\subsection*{Prompt optimization with few annotated papers}
\label{sec:prompt_optimization}

Using a seven-fold document-level leave-one-out design over the annotated molecular glue papers (Methods), we compared CAPO with the unoptimized seed prompt $P_0$ and GEPA~\cite{agrawal2025gepareflectivepromptevolution} under closely matched optimization budgets. CAPO increased the micro-averaged record-level $F_1$ score from 0.626 with $P_0$ to 0.980, compared with 0.907 for GEPA (Table~\ref{tab:prompt_optimization_comparison}). CAPO recovered all 98 ground-truth records (recall = 1.000) and achieved the highest macro-averaged record-level $F_1$, with substantially lower variation across held-out papers ($0.990 \pm 0.016$) than GEPA ($0.922 \pm 0.192$). Although its precision was slightly lower than that of GEPA (0.961 versus 0.976), CAPO achieved higher recall and overall record-level $F_1$ while preserving high field-level accuracy.

\begin{table}[t]
\centering
\small
\setlength{\tabcolsep}{2pt}
\caption{Prompt-optimization comparison under seven-fold document-level LOOCV.}\label{tab:prompt_optimization_comparison}
\begin{tabular}{l c ccc c c}
\toprule
\textbf{Method}
& \textbf{Opt. calls/fold}
& \multicolumn{3}{c}{\textbf{Micro Record}}
& \textbf{Macro Record}
& \textbf{Micro Field} \\
\cmidrule(lr){3-5}
& 
& P & R & F$_1$
& F$_1$
& F$_1$ \\
\midrule

Seed prompt ($P_0$)
& 0
& 0.785
& 0.520
& 0.626
& 0.805 $\pm$ 0.316
& 0.962 \\

GEPA
& 14.9
& \textbf{0.976}
& 0.847
& 0.907
& 0.922 $\pm$ 0.192
& 0.885 \\

CAPO
& 14.6
& 0.961
& \textbf{1.000}
& \textbf{0.980}
& \textbf{0.990 $\pm$ 0.016}
& \textbf{0.976} \\

\bottomrule
\end{tabular}
\end{table}

CAPO converged within three refinement rounds across the seven folds. Record-level precision improved from 0.83 to 0.98 and recall from 0.81 to nearly 1.00, both exceeding the acceptance threshold ($\tau = 0.9$; Supplementary Note 8). The largest improvements occurred in assay-record identification, whereas field-level performance was already above threshold at the first round and remained stable, indicating that refinement mainly corrected which assays were recognized rather than how their values were read. The resulting prompt revisions captured general extraction rules rather than paper-specific fixes, including aligning table values with the compound and cell line reported in the same row and retaining approximation and inequality qualifiers such as \textasciitilde, $<$ and $>$. Together, the increased record completeness and reduced cross-paper variation indicate that constrained, diagnosis-driven refinement is effective when only a few annotated papers are available. The selected prompt was then frozen and transferred to the PROTAC cohort by replacing ``molecular glue'' with ``PROTAC''.

\subsection*{Extraction accuracy against ground truth}
\label{sec:extraction_accuracy}

On the molecular glue cohort, the workflow achieved a record-level $F_1$ of 0.98 under both L1 and L2, with field-level $F_1$ scores of $0.98 \pm 0.01$ and $0.99 \pm 0.00$, respectively (Fig.~\ref{fig:accuracy_evaluation} and Table~\ref{tab:gt_accuracy}a). The same extraction prompt transferred to PROTACs by terminology substitution alone, maintaining high record-level performance with $F_1=0.93\pm0.02$ at L1 and $0.96\pm0.02$ at L2. Because the 15 PROTAC publications were not used at any stage of prompt optimization, they provide an independent test of the frozen prompt.

The same evaluation applied to the two curated databases shows a different pattern. Neither produced any record-level match under L1 ($F_1 = 0.00$), because most MolGlueDB and PROTAC-DB entries do not record the assay type or cell line required for the five-field key; this reflects a difference in the granularity at which the databases store assay context rather than an error in their reported values. At L2, MolGlueDB and PROTAC-DB recovered 32\% and 62\% of ground-truth records, yielding record-level $F_1$ scores of 0.43 and 0.69. The reduced MolGlueDB precision was driven by recruiter mismatches, in which database records listed ``Unknown'' where expert annotation specified ``SIAH1'' (Supplementary Note 4).
\begin{table}[ht]
\centering
\footnotesize
\caption{Extraction performance evaluated against ground truth on the MG and PROTAC cohorts. \textbf{a}, Record- and field-level performance of AutoTPD and the curated databases. L1 uses the 5-field key $(C,T,E,A,L)$, whereas L2 uses the
3-field key $(C,T,E)$, with assay and cell line scored as fields. Dashes indicate no record-level matches for field-level scoring.
\textbf{b}, Per-field extraction accuracy of AutoTPD at L1; bold values indicate the maximum possible score for that metric. Full field-level results are provided in Supplementary Note 2. AutoTPD values are mean (s.d.) over 3 independent runs for both cohorts. P, precision; R, recall; F$_1$, F$_1$ score.}
\label{tab:gt_accuracy}

\setlength{\tabcolsep}{2pt}
\renewcommand{\arraystretch}{1.0}
\begin{tabular}{lll ccc ccc}
\multicolumn{9}{@{}l}{\textbf{a}} \\
\toprule
 & & & \multicolumn{3}{c}{\textbf{Record-Level}} & \multicolumn{3}{c}{\textbf{Field-Level}} \\
\cmidrule(lr){4-6} \cmidrule(lr){7-9}
\textbf{Cohort} & \textbf{Source} & \textbf{Level} & P & R & F$_1$ & P & R & F$_1$ \\
\midrule
\multirow{4}{*}{MGs}
  & \multirow{2}{*}{AutoTPD}
    & L1 & 0.98 (0.01) & 0.98 (0.03) & 0.98 (0.02) & 0.99 (0.02) & 0.98 (0.01) & 0.98 (0.01) \\
  & & L2  & 0.98 (0.01) & 0.98 (0.03) & 0.98 (0.02) & 0.99 (0.01) & 0.99 (0.01) & 0.99 (0.00) \\
\cmidrule(lr){2-9}
  & \multirow{2}{*}{MolGlueDB}
    & L1 & 0.00 & 0.00 & 0.00 & --- & --- & --- \\
  & & L2  & 0.71 & 0.32 & 0.43 & 0.96 & 0.46 & 0.62 \\
\midrule
\multirow{4}{*}{PROTACs}
  & \multirow{2}{*}{AutoTPD}
    & L1 & 0.93 (0.02) & 0.95 (0.02) & 0.93 (0.02) & 0.96 (0.03) & 0.93 (0.01) & 0.93 (0.01) \\
  & & L2  & 0.96 (0.02) & 0.97 (0.01) & 0.96 (0.02) & 0.97 (0.02) & 0.95 (0.01) & 0.95 (0.01) \\
\cmidrule(lr){2-9}
  & \multirow{2}{*}{PROTAC-DB}
    & L1 & 0.00 & 0.00 & 0.00 & --- & --- & --- \\
  & & L2  & 0.90 & 0.62 & 0.69 & 0.92 & 0.78 & 0.84 \\
\bottomrule
\end{tabular}

\vspace{10pt}

\setlength{\tabcolsep}{5pt}
\begin{tabular}{l ccc ccc}
\multicolumn{7}{@{}l}{\textbf{b}} \\
\toprule
\textbf{Field}
& \multicolumn{3}{c}{\textbf{MG}}
& \multicolumn{3}{c}{\textbf{PROTAC}} \\
\cmidrule(lr){2-4}
\cmidrule(lr){5-7}
& P & R & F$_1$
& P & R & F$_1$ \\
\midrule
DC$_{50}$
& \textbf{1.00 (0.01)} & \textbf{1.00 (0.00)} & \textbf{1.00 (0.00)}
& 0.94 (0.01) & \textbf{1.00 (0.00)} & 0.97 (0.00) \\
DC$_{50}$ units
& \textbf{1.00 (0.00)} & \textbf{1.00 (0.00)} & \textbf{1.00 (0.00)}
& \textbf{1.00 (0.00)} & \textbf{1.00 (0.00)} & \textbf{1.00 (0.00)} \\
DC$_{50}$ hours
& 0.96 (0.07) & 0.95 (0.03) & 0.96 (0.02)
& 0.99 (0.01) & 0.91 (0.01) & 0.95 (0.01) \\
$D_{\mathrm{max}}$
& \textbf{1.00 (0.00)} & \textbf{1.00 (0.00)} & \textbf{1.00 (0.00)}
& 0.93 (0.05) & \textbf{1.00 (0.00)} & 0.96 (0.02) \\
$D_{\mathrm{max}}$ hours
& \textbf{1.00 (0.00)} & 0.98 (0.03) & 0.99 (0.01)
& 0.91 (0.07) & 0.65 (0.02) & 0.76 (0.03) \\
$D_{\mathrm{max}}$ conc.
& \textbf{1.00 (0.00)} & \textbf{1.00 (0.00)} & \textbf{1.00 (0.00)}
& 0.62 (0.22) & \textbf{1.00 (0.00)} & 0.77 (0.17) \\
\bottomrule
\end{tabular}
\end{table}

The same trend holds at the field level (Table~\ref{tab:gt_accuracy}b). The workflow extracted the core degradation measurements DC$_{50}$ and $D_{\max}$ with high accuracy in both cohorts ($F_1 \geq 0.96$). The curated databases, when their records did match, were accurate (MolGlueDB precision was 0.96 at L2, compared with 0.92 for PROTAC-DB), but their recall was low (0.46 and 0.78), indicating that curated values are reliable when present but incomplete in coverage. Most remaining errors in the workflow involved contextual fields rather than the core degradation values: in the PROTAC cohort, recall for $D_{\max}$ treatment time was 0.65 and precision for $D_{\max}$ concentration was 0.62, whereas all core endpoint values and units were at least 0.93. We return to the origin of these errors in the Discussion.

\subsection*{Robustness across LLM backbones}

To assess sensitivity to the underlying foundation model, we additionally evaluated the same workflow and CAPO-optimized prompt with DeepSeek-V3.1-Terminus and Gemini 2.5 Pro on the seven-paper molecular-glue cohort. All backbone comparisons used the same final expert-audited semantic cache, while the CAPO-optimized prompt, deterministic
matching rules, post-processing pipeline, and evaluation criteria were kept unchanged. The models were selected to span model families using versions appropriate to the study timeline, including one open-weight model. Across three independent runs, record-level F$_1$ remained high for GPT-5, DeepSeek-V3.1-Terminus, and Gemini 2.5 Pro (0.98, 0.99, and 0.93, respectively), with corresponding field-level F$_1$ values of 0.98, 0.94, and 0.96. Per-field analysis showed that endpoint values, units, and concentrations were extracted consistently across backbones, whereas treatment-time fields were more sensitive to backbone choice, particularly DC50 hours and Dmax hours (Supplementary Note 8). These results provide evidence that strong extraction performance is retained across different model families on the seven-paper MG benchmark, rather than being specific to GPT-5.

\subsection*{Database augmentation with AutoTPD}

\begin{figure}[!tbp]
    \centering
    \includegraphics[width=\linewidth]{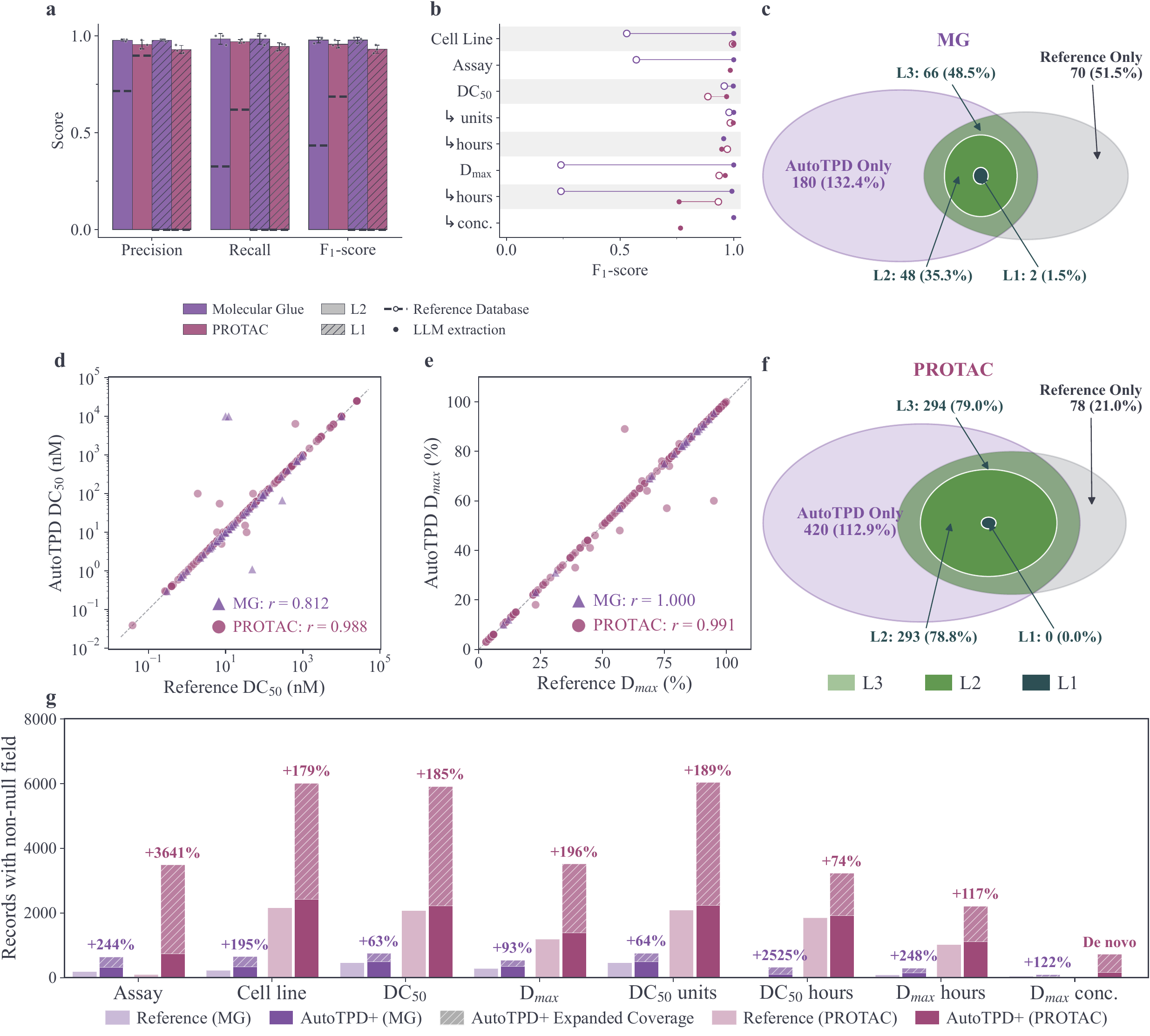}
    \caption{Evaluation of LLM extraction against expert annotations and reference databases, and database augmentation in AutoTPD+.
    \textbf{a}, Record-level precision, recall and F$_1$ scores for LLM extraction at L1 and L2. Bars show mean $\pm$ s.d. across seven MG leave-one-out folds or three PROTAC runs; dashed lines indicate the corresponding reference database records.
    \textbf{b}, Field-level F$_1$ scores at L2 for AutoTPD and reference database.
    \textbf{c,f}, Overlap between LLM-extracted records and reference database records at L1--L3 for MGs (\textbf{c}) and PROTACs (\textbf{f}). Percentages are relative to the corresponding reference database size; areas are not to scale.
    \textbf{d,e}, Agreement between matched reference-database and LLM-extracted DC$_{50}$ (\textbf{d}) and D$_{\max}$ (\textbf{e}) values. Dashed lines indicate identity, and Pearson's r is reported separately for MGs and PROTACs.
    \textbf{g}, Non-null field counts in the reference databases and AutoTPD+. Hatched segments indicate the additional coverage obtained by deploying the validated workflow beyond the benchmark cohorts. Percentages indicate changes in total AutoTPD+ coverage relative to the corresponding reference database; ``de novo'' denotes a field for which the reference database contains no values.}
    \label{fig:accuracy_evaluation}
\end{figure}

Within the benchmark cohorts of 70 MG and 141 PROTAC publications, we next evaluated the coverage gains provided by AutoTPD. For the baseline comparison, AutoTPD records were compared with cohort-matched, standardized records from the reference databases, restricted to source papers within the same benchmark cohorts rather than the full databases. For each cohort, we selected the run with the highest average record- and field-level $F_1$ as the released AutoTPD dataset. The MG and PROTAC AutoTPD datasets contained 246 and 714 records, representing net increases of 81\% and 92\% relative to the corresponding reference-database record sets of 136 and 372 records, respectively. Of these, 180 and 420 were AutoTPD-only records, respectively (Fig.~\ref{fig:accuracy_evaluation} c,f). After chemical-identifier enrichment, 196 of 246 MG records (79.7\%) and 556 of 714 PROTAC records (77.9\%) contained an RDKit-parseable canonical SMILES, yielding 752 structure enabled assay records overall (78.3\%). All 208 records without SMILES retained a source-reported compound name.

Stratified overlap analysis showed that AutoTPD–reference-database overlap increased as matching criteria were progressively relaxed from L1
(5 fields), through L2 (3 fields), to L3 (2 fields) in both cohorts (Fig.~\ref{fig:accuracy_evaluation}c), reflecting the limited assay-type and cell-line context recorded in the original databases. Matched records showed strong agreement in both cohorts, with Pearson correlations \(r = 0.81-0.99\) for DC$_{50}$ and \(r\geq 0.99\) for D$_{\max}$ (Fig.~\ref{fig:accuracy_evaluation}d \& e).

Manual review provided an independent estimate of the validity of records recovered by the LLM but absent from the source databases. We uniformly sampled without replacement 100 of 180 MG AutoTPD-only records and 40 of 420 PROTAC AutoTPD-only records using a fixed random seed of 42. Of these, 92/100 MG records and 33/40 PROTAC records were judged correct, corresponding to validity estimates of 92.0\% and 82.5\%, respectively. Conversely, reference-only records were dominated by figure-based evidence outside our text extraction scope (93\% and 65\%, respectively), with the remaining PROTAC cases split between missed SI tables and name mismatches. The same modality boundary also affected chemical structure resolution: most extracted assay records were associated with source reported compound names, while the corresponding molecular structures were presented only in chemical figures, precluding text-based SMILES recovery. Within the benchmark cohorts, AutoTPD substantially increased non-null field coverage relative to the standardized reference-database records: DC$_{50}$ coverage increased by 66\% for MGs and 68\% for PROTACs, D$_{\max}$ coverage by 89\% and 100\%, assay coverage by 498\% and 7,010\%, and cell-line coverage by 513\% and 98\%, respectively (Supplementary Note~3). These gains show that AutoTPD provides additional reliable records and substantially richer assay context, complementing the broader coverage of AutoTPD+ shown in Fig.~\ref{fig:accuracy_evaluation}g.

\subsection*{Large-scale deployment and construction of AutoTPD+}

Having validated the extraction workflow against expert annotations and standardized reference database records, we froze the extraction prompt and deployed the workflow across additional eligible PMC open-access molecular glue and PROTAC literature. After extraction, post-processing and release quality filtering, this deployment added 4,483 records from 598 publications beyond the AutoTPD benchmark, comprising 246 MG records from 44 publications and 4,237 PROTAC records from 554 publications.

The extracted records were integrated at the DOI level with standardized MolGlueDB and PROTAC-DB records. For publications processed by the LLM workflow, the extracted records replaced the corresponding source-database records; standardized source-database records were retained for publications outside the extraction scope. The resulting AutoTPD+ v1.0 release (October 2026) contains 856 MG and 6,766 PROTAC assay records, giving 7,622 records from 1,008 unique publications. AutoTPD+ uses stable record identifiers and is available through a public website (\url{https://yaochenr.github.io/AutoTPDplus/}), which provides release statistics, schema documentation, and direct downloads of the MG and PROTAC datasets.

\subsection*{Downstream utility for surrogate modeling}
We next evaluated whether the expanded AutoTPD+ affected downstream degradation-property modelling. The surrogate-model configurations, external test cohort and evaluation procedure were kept unchanged between the AutoTPD+ and PROTAC-DB comparisons. After excluding records overlapping with the external test set and records without valid molecular structures, the AutoTPD+ modelling pools contained 3,719 records for pDC$_{50}$ and 2,075 records for D$_{\max}$, compared with 1,937 and 1,031 PROTAC-DB records, respectively.

\begin{table}[ht]
\centering
\caption{Held-out performance of the same surrogate model configurations trained on AutoTPD+ or PROTAC-DB. Relative error reductions are calculated against PROTAC-DB. The 95\% confidence intervals were obtained from 2,000 paired DOI-cluster bootstrap resamples of the signed error difference
$\Delta=\mathrm{Error}_{\mathrm{PROTAC\mbox{-}DB}}-\mathrm{Error}_{\mathrm{AutoTPD+}}$.}
\label{tab:surrogate}
\begin{tabular}{llrrrr}
\toprule
Endpoint & Metric ($\downarrow$) & AutoTPD+ & PROTAC-DB & Error reduction & 95\% CI of $\Delta$ \\
\midrule
pDC$_{50}$ & RMSE & 1.267 & 1.391 & 8.9\%  & [$-0.024$, 0.255] \\
pDC$_{50}$ & MAE  & 1.012 & 1.166 & 13.1\% & [$-0.010$, 0.261] \\
D$_{\max}$ & RMSE & 20.140 & 23.675 & 14.9\% & [1.012, 3.958] \\
D$_{\max}$ & MAE  & 15.601 & 19.076 & 18.2\% & [$-0.398$, 4.784] \\
\bottomrule
\end{tabular}
\end{table}

Across all four endpoint--metric combinations, surrogate models using the best-performing TACK configurations~\cite{ribes2026tack} produced lower held-out prediction errors when trained on AutoTPD+ than when trained on PROTAC-DB (Table~\ref{tab:surrogate}). For pDC$_{50}$, RMSE decreased from 1.391 to 1.267 (8.9\%) and MAE from 1.166 to 1.012 (13.1\%). For D$_{\max}$, RMSE decreased from 23.675 to 20.140 (14.9\%) and MAE from 19.076 to 15.601 (18.2\%). The 95\% confidence interval excluded zero for D$_{\max}$ RMSE, whereas the intervals for both pDC$_{50}$ metrics and D$_{\max}$ MAE included zero. These results therefore indicate a consistent direction of relative improvement under the fixed modelling workflow, with the strongest statistical support observed for D$_{\max}$ RMSE.

\section*{Discussion}
\label{sec:discussion}

Our results suggest that the main bottleneck of manually curated TPD databases is coverage rather than accuracy: standardized reference-database records were reliable when present but recovered only 32\% (MolGlueDB) and 62\% (PROTAC-DB) of the ground-truth assays within the benchmark cohorts. Random audits of AutoTPD-only records found 92/100 MG and 33/40 PROTAC records to be correct, while many reference database only records relied on figure-based evidence outside our text-extraction scope. These findings support a shift from exhaustive manual transcription toward AI-assisted curation, in which domain experts provide targeted annotations and validate pipeline outputs. 

Because our primary goal is TPD database curation, the main benchmarks compare the workflow with existing databases and expert annotations. We nevertheless isolate the contribution of prompt refinement by comparing CAPO with the unoptimized seed prompt and GEPA under matched-budget document-level LOOCV~\citep{agrawal2025gepareflectivepromptevolution}. CAPO improves record completeness and cross-paper stability in this scarce-annotation setting. For TPD curation, the central question is whether automated extraction can recover records comparable in quality to existing curated resources. We address this through a domain-specific evaluation framework combining semantic and chemical-identifier matching, operator-aware endpoint comparison, and triangular comparison among LLM extracted records, standardized reference database records and expert annotated ground truth. Together, these components provide a reusable benchmark for future TPD extraction systems. The same two difficulties arise in other domains where structured records must be assembled from text, tables and supplementary files: adapting an extraction system from very few annotated documents, and validating it against a reference that is itself incomplete.

Several aspects of the evaluation warrant caution. Because the Champion Prompt was selected using held-out performance across the same seven molecular-glue folds on which accuracy was subsequently assessed, the molecular-glue results do not constitute a fully independent performance estimate and may be optimistic. The 15 PROTAC papers were not used during prompt optimization and therefore provide a more conservative test of transfer performance ($F_1 \geq 0.93$). The ground truth was also revised after comparison with the pipeline output, affecting approximately 5\% of records, although each revision was verified against the source article. Finally, the audits of newly recovered records sampled 100 of 180 molecular-glue records but only 40 of 420 PROTAC records, so the PROTAC validity estimate of 82.5\% carries a correspondingly wider uncertainty interval.

Although overall accuracy was high, the remaining errors were concentrated in contextual metadata and endpoint interpretation rather than the core DC$_{50}$ and D$_{\max}$ values. In the released AutoTPD molecular glue dataset, the only field-level errors were three missing DC$_{50}$ timepoints. In the released AutoTPD PROTAC dataset, 24 of the 29 record-level false negatives came from one paper, including 21 records associated with five entirely omitted compounds in a dense supplementary table.

Timepoint attribution was the most common challenge: our ground truth conservatively recorded timepoints only when explicitly linked to a specific measurement, whereas the pipeline sometimes inferred timepoints from context. These inferences appeared reasonable on manual inspection but were counted as errors under strict matching, explaining the lower scores for DC$_{50}$ hours in MGs and D$_{\max}$ hours in PROTACs. Differences reflected annotation scope, reporting conventions, and entity matching. For pDC$_{50}$-derived DC$_{50}$ values in PROTAC papers, the LLM, ground truth, and PROTAC-DB sometimes retained different decimal precision after conversion, creating small mismatches despite referring to the same measurement. For D$_{\max}$, our ground truth recorded D$_{\max}$ only when it was explicitly reported as such, whereas the pipeline sometimes extracted related quantities such as maximum degradation percentages or single-concentration degradation values into the D$_{\max}$ field. This was the dominant error among the invalid AutoTPD-PROTAC-only records identified during manual review, reflecting reporting ambiguity in the literature and the difficulty of manual curation at the scale of PROTAC-DB. Less frequent errors included EC$_{50}$--DC$_{50}$ confusion, where related potency metrics were reported in similar contexts, and compound-identity mismatches caused by inconsistent names or stereochemical identifiers. Representative failure cases are provided in Supplementary Note 10.

Beyond extraction accuracy, the value of AutoTPD lies not only in its increased record coverage (81\% for molecular glues and 92\% for PROTACs relative to the corresponding standardized reference records), but also in the richer assay context it provides. Treatment times, concentrations, assay types and cell lines, which are frequently missing from manually curated databases, are recovered alongside the core DC$_{50}$ and D$_{\max}$ endpoints. These fields matter because degradation activity is not a context-free compound property; comparing potency across studies and training condition-aware models both depend on the assay configuration under which each value was measured. By retaining weak and inactive observations alongside potent compounds, the workflow also broadens the activity distribution available for modeling. 

Having validated the workflow on the benchmark cohorts, we froze the extraction prompt and deployed it across additional eligible PMC open-access literature. This deployment added 4,483 records from 598 publications and, after integration with standardized reference-database records, produced AutoTPD+ v1.0, containing 7,622 records from 1,008 publications. Separating the frozen AutoTPD benchmark from the versioned AutoTPD+ living database enables reproducible evaluation while allowing the resource to be updated as new literature becomes available. Under a fixed modelling workflow and external test cohort, models trained on AutoTPD+ produced 8.9--18.2\% lower held-out error point estimates than the same models trained on PROTAC-DB. These results therefore support a consistent direction of relative improvement under the controlled comparison, rather than establishing general predictive superiority.

The main limitation of the workflow is its restriction to text and table modalities. Some records lacked SMILES because compound structures were reported only in figures, outside the workflow’s text-and-table scope. Because many reference database only records were figure-supported (93\% for molecular glues, 65\% for PROTACs), the expansion reflects augmentation of text- and table-accessible evidence rather than complete literature coverage; vision-language extraction from plots, blots and other figures is the most direct path to improving coverage. Additional scope constraints include the small but audited ground-truth set (7 molecular glue and 15 PROTAC papers), restriction to open-access PMC content, and evaluation focused on degradation endpoints. The surrogate-model comparison was further limited by the number of independent DOI clusters with evaluable endpoints, particularly for D$_{\max}$, and was designed to assess the relative effect of the training dataset under a fixed modelling workflow rather than absolute model performance. Future work should expand the expert annotations and independent held-out cohorts to tighten confidence intervals and extend the workflow to figures, ubiquitination, permeability, metabolism and other TPD-relevant measurements. Even within these bounds, the workflow recovers a substantial body of machine-readable assay data that existing manual curation has not reached, and its components are not specific to targeted protein degradation. We release the frozen AutoTPD benchmark and AutoTPD+ v1.0, together with a versioned update workflow, prompts, expert annotations and evaluation framework. AutoTPD+ v1.0 is accessible through a public website, and the accompanying workflow is configured for monthly literature searches and future versioned releases. Together, the database and workflow provide a resource for the TPD community and a template for AI-assisted curation in scientific domains where structured data remain scattered across the literature.

\section*{Methods}\label{sec:methods}

\subsection*{Data acquisition and pre-processing}
Our pipeline takes as input full-text scientific articles and their supplementary materials. For benchmark evaluation, we constructed three datasets supporting a triangular comparison: standardized reference database records derived from existing manually curated databases, LLM-extracted records produced by the workflow, and independently expert annotated ground-truth records. Here, ``reference-database records'' refers to the cohort-matched, standardized records derived from MolGlueDB and PROTAC-DB, whereas the ``baseline comparison'' refers to the comparison of AutoTPD with these records. Expert annotations are treated separately as ground truth. Comparing LLM-extracted records with ground truth measures extraction accuracy; comparing standardized reference database records with ground truth audits the coverage of the existing databases; and comparing LLM-extracted records with reference database records quantifies how much existing information the workflow recovers and how many additional records it identifies. We describe the construction of these datasets below.

\subsection*{Reference dataset construction and standardization}
\label{sec:baseline_standardization}

We sourced our study cohorts from two established TPD databases: PROTAC-DB 3.0 (released February 2025) and MolGlueDB (released July 2025), both accessed in August 2025~\citep{ge2025protac,wang2026molgluedb}. Using the listed endpoint-source DOIs, we retrieved the full-text articles from PubMed Central (PMC), retaining those with accessible open-access content ($N = 141$ PROTAC and $N = 70$ MG papers;  Supplementary Note 5). Because both databases store DC$_{50}$ and $D_{\max}$ values in free-text fields, often with mixed units and multiple assays in a single entry, we standardized the records by parsing free-text values, normalizing units, and splitting merged entries into one degradation assay record per row; entries lacking both DC$_{50}$ and $D_{\max}$ were excluded. The resulting standardized records were treated as the \textit{reference database} dataset for benchmark comparison. A stratified subset of the source publications was independently re-annotated to construct the expert ground truth.

\subsection*{Expert ground truth annotation}
\label{subsubsec:ground truth}

We constructed an independent ground-truth dataset for evaluation by stratified random sampling of approximately 10\% of papers from each cohort (PROTAC: 15 papers; MolGlue: 7 papers), drawing from multiple journals to span variation in article structure and reporting style. Details are provided in Supplementary Note 1. Each record was independently reviewed by two annotators using the full text and supplementary materials, restricted to textual and tabular content to match the scope of the LLM pipeline. Annotations followed a conservative policy: a value was recorded only when explicitly supported by the source document, and values that could be inferred but were not directly stated were omitted. During comparison against pipeline predictions, we identified a number of cases in which the LLM extraction disagreed with our initial annotations and, on manual re-inspection of the source articles, found the pipeline to be correct; we used these flagged disagreements to systematically audit the ground-truth set, and revised the annotations in a second pass. Reported ground-truth metrics use this revised version.

\subsection*{LLM extraction workflow}
\label{sec:CAPO}

We developed an expert-in-the-loop LLM workflow to extract structured TPD assay records from full-text articles and supplementary materials without model fine-tuning or large-scale annotated corpora. Because expert annotation of full scientific papers is costly, the workflow uses a lightweight cross-validated prompt-refinement component, CAPO (Cross-validated Agentic Prompt Optimization), to adapt extraction instructions from a small set of expert-annotated papers. The CAPO-optimized MG prompt was frozen, and a PROTAC prompt was derived from it by terminology substitution alone. The two prompts were then applied to their respective benchmark cohorts, followed by domain specific post-processing and evaluation.

As shown in Fig.~\ref{fig:pipeline_overview}, the prompt-refinement module operates as a cross-validated dual-loop procedure. The inner loop performs paper-level extraction, where the Extractor agent processes each article through multi-stage reasoning. The outer loop iteratively refines the extraction prompt through an Assess--Diagnose--Refine cycle. In each round, the current prompt is applied to the optimization batch and evaluated against ground truth. If the batch-averaged precision and recall meet the acceptance threshold ($\tau = 0.9$), refinement converges and the prompt is finalized for that fold. Otherwise, the Analyst agent diagnoses general patterns in below-threshold papers, and the Refiner agent generates an updated prompt for the next round. After benchmark validation, both prompts were frozen and used without further optimization for literature-scale deployment and subsequent monthly updates.

\subsection*{Iterative prompt optimization logic}

The optimization runs as $J$ independent LOOCV folds: in each, $J-1$ annotated papers form the optimization batch and one paper is held out for validation, yielding one candidate prompt per fold and exposing the Refiner agent to six of the seven annotated papers (86\%) in every fold. Each fold is initialized from the same seed prompt $P_0$, which uses a four-section template (Goal, Required Fields, Extraction Principles, Output Format) but leaves Extraction Principles empty so that the optimization process must discover the relevant extraction rules from observed document-level extraction failures rather than from manual trial-and-error. Within a fold, papers in the batch that fall below the local acceptance threshold $\tau$ are routed to the Analyst agent for error diagnosis; the resulting paper-level diagnoses are aggregated across the optimization set into a diagnostic summary used by the Refiner to update the prompt for the next round (Supplementary Note 8). Thus, prompt updates are driven by recurring extraction failures identified across documents rather than by unrestricted generation or search over candidate prompts. To limit overfitting to the small annotation set, revisions are restricted to general extraction principles: required fields and the output schema remain fixed, paper-specific entities cannot be hard-coded, and each refinement step is limited to at most two added or modified sentences (Supplementary Note 8). After all $J$ folds converge, we obtain a set of candidate prompts $\mathcal{P}_{\mathrm{cand}} = \{P_1, \dots, P_J\}$ and select the final prompt $P^*$ (also referred to as the Champion Prompt) as the one achieving the highest $F_1$-score on its held-out paper, $P^* = \arg\max_{P_i \in \mathcal{P}_{\mathrm{cand}}} F_1(P_i)$. We used $F_1$ to balance precision and recall, avoiding selection based solely on either missed records or false-positive control.

\subsection*{Agentic architecture configuration}
\label{sec:agent_config}

The primary experiments and CAPO prompt optimization use three GPT-5 agents (OpenRouter identifier: \texttt{openai/gpt-5}): an Extractor, an Analyst, and a Refiner, together with a separate semantic matching module. The Extractor runs at temperature 0.0 to minimize sampling variability; however, temperature-zero inference through the hosted API was not assumed to be fully deterministic. Extraction was therefore repeated in three independent runs, and variability across runs is reported as the mean $\pm$ s.d. The Analyst runs at temperature 0.1 for stable error diagnosis, whereas the Refiner runs at 0.7 to allow diversity during prompt revision. Prompt templates are provided in Supplementary Note 7. Each extraction pass requested a pass-specific JSON response using prompt-level schema instructions and was parsed post hoc. The \textbf{Extractor} processes each paper in four phases: initial extraction from the main text; sentence-level source verification; integration of supplementary materials; and a final main-text review to complete missing chemical identifiers without changing assay measurements. Conversation state persists within a paper and is reset between papers. The \textbf{Analyst} performs root-cause analysis on the papers that fall below the acceptance threshold, receiving the source document, categorized false negatives and false positives, the ground truth, and prior semantic-matching decisions (which prevent valid synonym matches from being misdiagnosed as errors). It emits a concise diagnostic summary per failed document. The \textbf{Refiner} aggregates the per-paper diagnoses into a batch-level update, constrained to extraction principles only (output schema and required fields are frozen). This focuses each refinement on common error patterns rather than paper-specific fixes. The \textbf{Semantic Matching Module} is used only during evaluation to resolve non-exact terminology matches for degradation targets, recruiters, assay types, and cell lines. For a previously unseen pair, a secondary LLM using the evaluated backbone at temperature zero makes a constrained yes/no decision. Compound identities are matched separately using deterministic name and chemical-identifier rules with a Standard InChIKey fallback, while quantitative fields are compared using deterministic operator-aware numerical criteria. Semantic-matching decisions are stored in a single shared cache that is reused across documents, extraction runs, and backbone evaluations. All non-exact decisions in the final cache are expert reviewed and corrected where necessary, and all reported backbone comparisons use this same final expert-audited cache. The module does not participate in record extraction or prompt optimization; implementation and reliability-audit details are provided in Supplementary Note 9.

\subsection*{Full-scale dataset construction and update workflow}

The frozen AutoTPD benchmark datasets were generated from LLM extracted records for the database sourced PMC cohorts through normalization, deduplication, identifier enrichment and standardization. The resulting benchmark contained 246 MG and 714 PROTAC assay records and was used for extraction accuracy and database augmentation analyses.

To extend coverage beyond these cohorts, we developed a PMC literature mining and screening workflow that runs predefined MG and PROTAC searches, consolidates results into a DOI-, PMID- and PMCID-resolved publication registry, screens articles for relevant degradation evidence, and retrieves eligible open-access full texts and supplementary files. Full search queries, screening criteria and implementation details are provided in Supplementary Note~5. Eligible publications were processed using the frozen extraction prompts and the same post-processing workflow used for AutoTPD. This deployment added 4,483 records from 598 publications beyond the benchmark. DOI-level integration of the extracted records with standardized reference-database records produced AutoTPD+ v1.0, containing 856 MG and 6,766 PROTAC records, giving 7,622 records from 1,008 publications. The update workflow is configured to rerun the PMC searches monthly. Newly eligible publications will be incorporated into future versioned releases using the frozen extraction workflow, automated quality checks and expert review.

\subsection*{Evaluation framework and metrics}
\label{subsec: evaluation metrics}

Each predicted record is compared with the reference at one of three nested matching levels. L1 uses the full 5-element key $R=\langle C,T,E,A,L\rangle$, where $C$ denotes compound identity resolved within the source DOI using the normalized compound name or label, with a Standard InChIKey fallback when available (SMILES was not required), $T$ the degradation target, $E$ the recruiter, $A$ the assay type, and $L$ the cell line. L2 uses the 3-element key $R=\langle C,T,E\rangle$ and treats $A$ and $L$ as field-level quantities. For large-scale LLM–reference-database discrepancy analysis, we additionally use L3, defined by $R=\langle C,T\rangle$, to measure compound--target overlap independent of recruiter, assay type, or cell line. For ground-truth evaluation, we report L1 and L2, both of which retain recruiter identity. L3 is reserved for large-scale AutoTPD–reference-database overlap analysis based only on compound--target identity.

At each level, evaluation is hierarchical. \textit{Record-level} evaluation determines whether the pipeline correctly identifies a degradation assay, using compound-name matching with a Standard InChIKey fallback and two-round conflict resolution (Supplementary Note~6). \textit{Field-level} evaluation, applied only to record-level true positives (TPs), assesses whether the quantitative columns within a matched row are correct under operator-aware comparisons that handle unit equivalences (e.g., nmol/L vs. nM), inequalities ($<$, $>$), and numerical ranges (Supplementary Note 6). At both levels we report precision, recall, and $F_1$, computed under standard TP/FP/FN definitions; true negatives are excluded from metric calculations. Record-level micro-averaged $F_1$ serves as the primary indicator of extraction performance (Supplementary Note~6).

\subsection*{Comparative analysis framework}

We evaluated the workflow using a triangular comparison of standardized reference database records ($D_{\mathrm{ref}}$), LLM extracted records ($D_{\mathrm{LLM}}$), and independently annotated ground truth ($D_{\mathrm{gt}}$). The three comparisons measure complementary quantities: $D_{\mathrm{LLM}}$ versus $D_{\mathrm{gt}}$ measures extraction accuracy, $D_{\mathrm{ref}}$ versus $D_{\mathrm{gt}}$ audits reference database coverage, and $D_{\mathrm{LLM}}$ versus $D_{\mathrm{ref}}$ quantifies database augmentation. For the last analysis, records were classified as shared, reference database-only or LLM extraction-only.

To assess the validity of newly recovered records, we drew simple random samples without replacement from the complete AutoTPD-only record pools using a fixed random seed of 42, with each record having equal sampling probability. We manually reviewed 100 of 180 MG records and 40 of 420 PROTAC records. Sampling was performed at the record level without stratification and was distinct from the stratified paper-level sampling used to construct the ground-truth cohorts. For reference-only records, all 70 MG records and a simple random sample of 20 PROTAC records, drawn using the same procedure, were manually reviewed.

\subsection*{Downstream surrogate-model evaluation}

To assess the relative effect of the expanded training data, we compared models trained on AutoTPD+ with models trained on PROTAC-DB using the official TACK implementation~\citep{ribes2026tack}. The task specific feature representations, model configurations and training procedure were fixed across the two corpora. Following TACK, pDC$_{50}$ was modelled using a multilayer perceptron and D$_{\max}$ using XGBoost. The external test cohort comprised independently annotated records from 15 PROTAC publications and was used only for final evaluation.

To prevent information leakage, we excluded training records whose source DOI or canonical-SMILES compound identity overlapped with the test cohort, as well as records lacking an RDKit parseable SMILES or the relevant endpoint. The resulting AutoTPD+ and PROTAC-DB training pools contained 3,719 and 1,937 pDC$_{50}$ records, representing 1,805 and 1,271 unique compounds, respectively. The corresponding D$_{\max}$ pools contained 2,075 and 1,031 records, representing 1,099 and 659 compounds.

For each corpus and endpoint, we trained a 25-model ensemble using compound-grouped approximately 90:10 training--validation splits, with validation used only for early stopping. Ensemble predictions were obtained by averaging across models. DC$_{50}$ values were converted to
$p\mathrm{DC}_{50}=9-\log_{10}(\mathrm{DC}_{50}\,[\mathrm{nM}])$.

The primary evaluation used exact and approximate ($\sim$) measurements and reported RMSE and MAE. Inequality-censored measurements were evaluated separately using an interval-aware procedure; its definition and results are reported in Supplementary Note~11. Differences between AutoTPD+ and PROTAC-DB were evaluated using 2,000 paired DOI-cluster bootstrap resamples of the signed error difference, with a fixed random seed. Percentage error reduction was calculated relative to the PROTAC-DB error.

\subsection*{Use of large language models}
GPT-5, DeepSeek-V3.1-Terminus and Gemini 2.5 Pro were used as research tools for data extraction, error diagnosis and prompt refinement, as described above; no model was fine-tuned. The authors also used large language models to improve the language and readability of text in the manuscript.

\backmatter

\bmhead{Data availability}
The AutoTPD and AutoTPD+ datasets generated in this study, together with the standardized reference-database records, expert-annotated ground-truth records and manual-audit results used in the evaluation, are publicly available through the project website at \url{https://yaochenr.github.io/AutoTPDplus/} and Hugging Face at \url{https://huggingface.co/datasets/ailab-bio/AutoTPDplus}. The source publication DOIs used to construct the molecular glue and PROTAC cohorts are included in the released datasets. Full-text articles and supplementary materials are not redistributed and remain available from their original publishers, subject to the applicable licenses.

\bmhead{Code availability}
The source code used for full-text and supplementary-material parsing, CAPO prompt optimization, LLM-based extraction, post-processing, record matching and evaluation is publicly available at \url{https://github.com/yaochenr/AutoTPDplus}. The repository includes the extraction and evaluation code, prompt templates, configuration files, analysis notebooks and a Pixi environment lockfile for reproducing the reported analyses from the released datasets. The source code is released under the MIT License.

\bmhead{Acknowledgements}

The authors thank Dr. Richard Beckmann, Nils Dunlop, Sofia Larsson, Pablo Martínez Crespo, and Stefano Ribes, for their support on the manual annotation and audit of the ground-truth dataset. We also thank Stefano Ribes for his insights on the statistical validation framework for the downstream use case. We are also grateful to Dr.~Kevin Maik Jablonka and Mara Schilling-Wilhelmi for valuable feedback on this work. 

\section*{Author contributions}
\textbf{YR}: Conceptualization, Methodology, Software, Data curation, Formal analysis, Investigation, Validation, Visualization, Writing -- original draft, Writing -- review \& editing.
\textbf{FJ}: Data curation, Validation, Writing -- review \& editing.
\textbf{NMAK}: Methodology, Writing -- review \& editing.
\textbf{RM}: Conceptualization, Methodology, Data curation, Validation, Supervision, Project administration, Funding acquisition, Resources, Writing -- review \& editing.

\section*{Competing interests}
The authors declare no competing interests.

\section*{Funding}

YR acknowledges funding from the Swedish Research Council (VR); FJ and RM acknowledge funding and support provided by the Wallenberg AI, Autonomous Systems, and Software Program (WASP), supported by the Knut and Alice Wallenberg Foundation. The computations and data storage were enabled by resources provided by Chalmers e-Commons and by the National Academic Infrastructure for Supercomputing in Sweden (NAISS), partially funded by the Swedish Research Council through grant agreement no. 2022-06725.

\nocite{chen2025development,huckvale2022improved,nishiguchi2021identification,hassan2024exploration,li2024novel,liu2023rational,mercer2024continuous,chan2018impact,alabi2021mutant,son2023chemical,macabuag2022developing,schroder2024dcaf1,ng2023heterobifunctional,testa2020structure,kofink2022selective,gabizon2020efficient,li2023discovery,xiao2021proteolysis,yu2021selective,lv2021development,farnaby2019baf,steinebach2023leveraging,paruchuri_marker_2026,python_docx}

\bibliography{sn-bibliography}

\clearpage
\includepdf[pages=-]{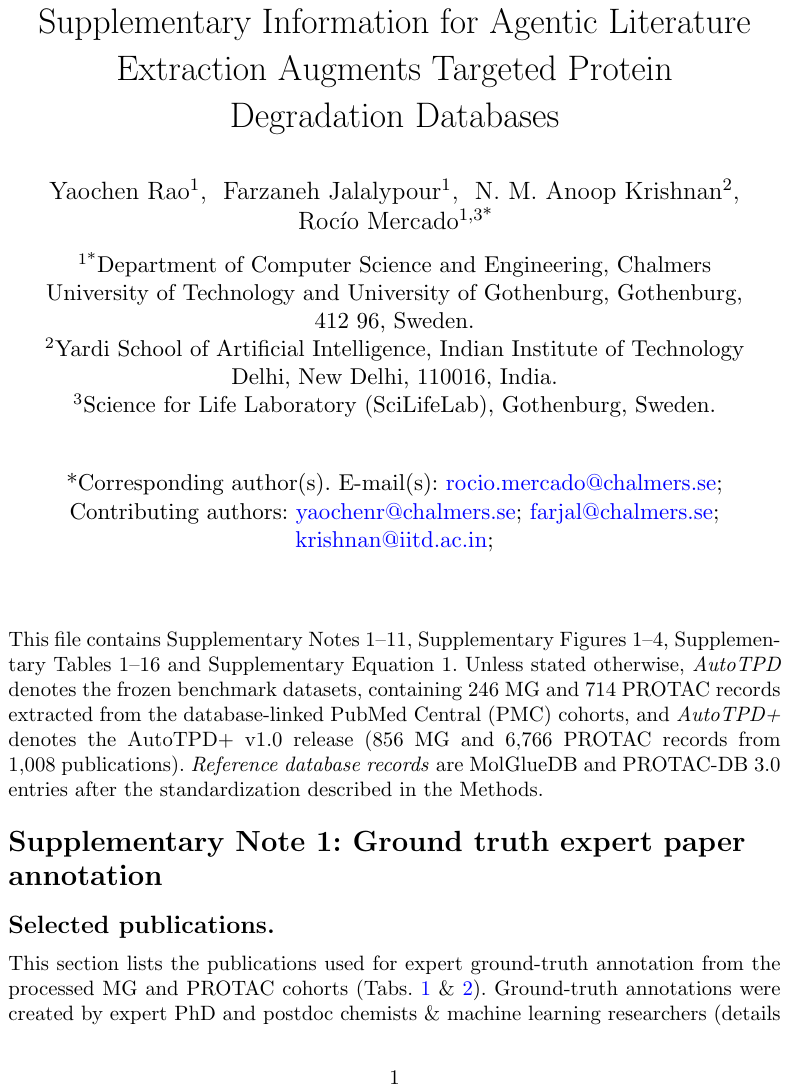}

\end{document}